\documentclass[pra,twocolumn,showpacs,amsmath]{revtex4}

\newcommand{\subsize} {\scriptsize}

\newcommand{\sub}[1] {_\textrm{\subsize #1}}

\newcommand{\etalcite}[2] {#1 \emph{et al.}~\cite{#2}}

\newcommand{\reffig}   [1]{fig.}
\newcommand{\reffigs}  [1]{figs.}
\newcommand{\Reffig}   [1]{Fig.}

\newcommand{\smartfig} [1]{\reffig.~\ref{#1}}
\newcommand{\smartfigs}[1]{\reffigs.~\ref{#1}}
\newcommand{\Smartfig} [1]{\Reffig.~\ref{#1}}

\newcommand{\refeq}[1] {eq.}
\newcommand{\refeqs}[1] {eqs.}
\newcommand{\Refeq}[1] {Eq.}
\newcommand{\Refeqs}[1] {Eqs.}

\newcommand{\smarteq}[1] {\refeq.~(\ref{#1})}
\newcommand{\smarteqs}[1] {\refeqs.~(\ref{#1})}
\newcommand{\Smarteqs}[1] {\Refeqs.~(\ref{#1})}
\newcommand{\Smarteq}[1] {\Refeq.~(\ref{#1})}
\newcommand{\twoeqs}[2] {\refeqs.~(\ref{#1}) and (\ref{#2})}

\usepackage[dvips]{graphicx}   
\usepackage{epsfig}

\begin{document}

\title{Stability, effective dimensions, and interactions for bosons in
  deformed fields}

\author{O.~S\o rensen}
\author{D.~V.~Fedorov}
\author{A.~S.~Jensen}

\affiliation{Department of Physics and Astronomy, University of
  Aarhus, DK-8000 Aarhus C, Denmark}

\date{\today}

\begin{abstract} 
  The hyperspherical adiabatic method is used to derive stability
  criteria for Bose-Einstein condensates in deformed external fields.
  An analytical approximation is obtained.  For constant volume the
  highest stability is found for spherical traps.  Analytical
  approximations to the stability criterion with and without zero
  point motion are derived.  Extreme geometries of the field
  effectively confine the system to dimensions lower than three.  As a
  function of deformation we compute the dimension to vary
  continuously between one and three.  We derive a dimension-dependent
  effective radial Hamiltonian and investigate one choice of an
  effective interaction in the deformed case.
\end{abstract}


\maketitle

\section{Introduction}

Bose-Einstein condensation is routinely achieved in a number of
laboratories \cite{sac99,ste99,gor01b,rob01}, see further descriptions
in the recent monographs \cite{pet01,pit03}.  The techniques involve
cooling and trapping of atomic gases in external laser fields and
magnetic fields.  These traps are in practice of cylindrical geometry
\cite{rob01,don01,cla03}.  For $N$ attractive atoms the stability
criterion is experimentally established to be $N |a_s| / b\sub t <
0.55$ \cite{cla03}, where $a_s$ is the scattering length and $b\sub
t\equiv \sqrt{\hbar/(m \omega)}$ is the relevant length scale of the
harmonic trap of geometric average frequency $\omega
\equiv\sqrt[3]{\omega_x\omega_y\omega_z}$.  A reduction from three
dimensions to effectively one or two dimensions was observed
experimentally \cite{gor01b} in the limit when the interaction energy
is small compared to the level spacing in the tightly-confining
dimension.  Experiments with continuous variation of the trap geometry
from three to either one or two effective dimensions \cite{gor01b},
with a two-dimensional structure \cite{gre01,ryc03}, and an effective
one-dimensional geometry \cite{tol03} request a corresponding
theoretical description.

Theoretical interpretations and the underlying analyses are frequently
based on model assumptions of spherical symmetry \cite{boh98,adh02b}.
Confinement to lower dimensions can also be studied directly without
the three-dimensional starting point \cite{kim00}.  This has been done
with a variational calculation in Gross-Pitaevskii equation (GPE)
\cite{bay96} and more recently in the GPE with variational
dimensionality \cite{mck02b}.  Also effects on stability of deformed
external fields have been investigated by use of the GPE formulation
\cite{bay96,gam01,adh01c}.  Extreme deformations could result in
effective one-dimensional or two-dimensional systems which can be
described by effective interactions of corresponding discrete lower
dimensions \cite{ols98,pet00,pet01b,lee02}.

In the present article we rewrite the hyperspherical formulation from
reference \cite{sor02b} to account for a general deformation of the
external field.  Since two-body correlations are not yet included in
the wave function, this hyperspherical approach resembles a mean-field
treatment.  We investigate the stability criterion in section
\ref{sec:stabilitydeformed}.  Section \ref{sec:effdim} contains an
approach to an effective dimension which depends on the deformation of
the external field.  Since the interactions are presently not included
in this effective dimension ($d$), we therefore in section
\ref{sec:effint} introduce them on top of the derived $d$-dimensional
Hamiltonian.  Although the choice of interactions is not unique, we
can with some guess obtain an alternative stability criterion and
subsequently interpret the results in terms of a deformation-dependent
coupling strength, which is finally compared with known results.

\section{Hyperspherical description}

A combination of magnetic fields results in an effective trapping
potential, which can be described as the deformed harmonic oscillator
potential $V\sub{trap}$ acting on all the identical particles of mass
$m$
\begin{eqnarray}
  V\sub{trap}(\boldsymbol r_i)
  =
  \frac{1}{2}m
  \big(\omega_x^2 x_{i}^2
  +\omega_y^2y_{i}^2
  +\omega_z^2z_{i}^2\big)
  \;,
\end{eqnarray}
where the position of the $i$th particle is $\vec r_i =
(x_{i},y_{i},z_{i})$ and the frequencies along the coordinate
directions $q=x,y,z$ are denoted $\omega_q$.  The hyperradius $\rho$
is the principal coordinate, which is separated into the components
$\rho_x$, $\rho_y$, and $\rho_z$ along the different axes, i.e.
\begin{eqnarray}
  &&
  \rho^2
  =
  \frac{1}{N}
  \sum_{i<j}^Nr_{ij}^2 = \rho_x^2+\rho_y^2+\rho_z^2
  \equiv \rho_\perp^2 + \rho_z^2
  \;,
\end{eqnarray}
where $\boldsymbol r_{ij} = \boldsymbol r_j - \boldsymbol r_i$.  In
the center-of-mass system the remaining coordinates are given as
angles collectively denoted by $\Omega$.

An application here of the method presented in reference \cite{sor02b}
is to assume a relative wave function as a sum of two-particle
components.  In the case of a spherical trapping field each two-body
component only needs dependence on $\rho$ and the two-body distance
$r_{ij}=\sqrt2\rho\sin\alpha_{ij}$ through an angle $\alpha_{ij}$.
For a deformed external field it also needs dependence on the angle
$\vartheta_{ij}$ between the interatomic vector $\boldsymbol r_{ij}$
and the axis of the external field.  The two-body component should in
the cylindrical case then be on the form
$\phi(\rho,\alpha_{ij},\vartheta_{ij})$, which would lead to an
angular equation in the two variables $\alpha_{12}$ and
$\vartheta_{12}$ with complicated integrals.  We shall here restrict
ourselves to a wave function which is independent of hyperangles.
This is expected to give a fair description for dilute systems where
the large distances average out all the directional dependence
\cite{sor03b}.  This is in contradiction with our conclusions in
ref.~\cite{sor01} for repulsive interactions.  The mistake was later
corrected, see the treatment of attractive interactions in
ref.~\cite{sor02b}.

Thus, we neglect correlations in analogy to a mean-field treatment, so
in the dilute limit the hyperangular average of the relative
Hamiltonian is
\begin{eqnarray}
  &&
  \langle \hat H\rangle_\Omega
  \;\to\;
  \hat H
  =
  \hat H_{x} + \hat H_{y} + \hat H_{z}
  +  \hat V
  \;,\\
  &&
  \hat V = \sum_{i<j}^N
  \big\langle V_{ij}  \big\rangle_\Omega
  \;,
  \\
  &&
  \frac{2m\hat H_q}{\hbar^2}
  =
  -\frac{1}{\rho_q^{d(N-1)-1}}
  \frac{\partial}{\partial\rho_q}
  \rho_q^{d(N-1)-1}
  \frac{\partial}{\partial\rho_q}
  +\frac{\rho_q^2}{b_q^4}
  \label{eq:gend1}
  \;,
\end{eqnarray}
where $d=1$ and $b_q^2 \equiv \hbar/(m\omega_q)$ for $q=x,y,z$. The
interactions $V_{ij}$ are averaged over all angles $\Omega$, which for
the zero-range interaction $4\pi\hbar^2a_s\delta(\boldsymbol
r_{ij})/m$  for $N\gg1$ yields
\begin{eqnarray}
  \hat V
  =
  \frac{4\pi\hbar^2 a_s}{m} \sum_{i<j}^N  
  \big\langle \delta(\boldsymbol r_{ij}) \big\rangle_\Omega 
  = 
  \frac{\hbar^2}{2m}
  \frac{1}{2\sqrt{\pi}}
  N^{7/2}
  \frac{a_s}{\rho_x \rho_y \rho_z}  
  \;.
\end{eqnarray}
If we replace as $\rho_x=\rho_y=\rho_z=\rho/\sqrt3$, this is identical
to the average of the zero-range interaction in the spherically
symmetric case \cite{sor02b}.

We define the following dimensionless coordinates and parameters:
\begin{eqnarray}
  &&
  x\equiv \frac{\rho_x}{b_x} \sqrt{\frac{2}{N}}
  \;, \qquad
  y\equiv \frac{\rho_y}{b_y} \sqrt{\frac{2}{N}}
  \;, \qquad
  z\equiv \frac{\rho_z}{b_z} \sqrt{\frac{2}{N}}
  \;,\quad
  \\
  &&
  \beta \equiv \frac{b_x^2 + b_y^2}{2 b_z^2}
  \;, \qquad
  \gamma \equiv \frac{b_x^2 - b_y^2}{2 b_z^2}
  \;, \qquad
  s\equiv \frac{Na_s}{b\sub t}
  \;, \\
  &&
  b\sub t^3\equiv b_x b_y b_z
  \;.\;
\end{eqnarray}
The deformation along the different axes is then described by $\beta$
and $\gamma$, and $s$ is the effective interaction strength.  The
Schr\"odinger equation $\hat H F(\rho_x,\rho_y,\rho_z) =
EF(\rho_x,\rho_y,\rho_z)$ is rewritten with the transformation
\begin{eqnarray}
  f(x,y,z)
  \propto
  (xyz)^{(N-2)/2} F(\rho_x,\rho_y,\rho_z)
  \label{eq:fFtrans}
\end{eqnarray}
in order to avoid first derivatives.  We then obtain
\begin{eqnarray}
  &&
  \bigg[
  -\frac{1}{\beta+\gamma}\frac{\partial^2}{\partial x^2}
  -\frac{1}{\beta-\gamma}\frac{\partial^2}{\partial y^2}
  -\frac{\partial^2}{\partial z^2}
  \nonumber\\
  &&
  \qquad
  +\frac{N^2u(x,y,z)-\varepsilon}{2\sqrt[3]{\beta^2-\gamma^2}}
  \bigg]
  f(x,y,z)=0
  \label{eq:gend16}
  \;,\\
  &&
  u(x,y,z)
  =
  \frac12\sqrt[3]{\beta^2-\gamma^2}\bigg[
  \frac{1}{\beta+\gamma}\Big(x^2+\frac{1}{x^2}\Big)
  \nonumber\\
  &&
  \qquad
  +\frac{1}{\beta-\gamma}\Big(y^2+\frac{1}{y^2}\Big)
  +z^2+\frac{1}{z^2}\bigg]
  +\sqrt{\frac{2}{\pi}}
  \frac{s}{xyz}
  \label{eq:gend17}
  \;,\quad
\end{eqnarray}
where $\varepsilon \equiv 2NE/(\hbar\omega)$.  Without interaction,
i.e.~$a_s=0$, the ground-state solution is
\begin{eqnarray}  \label{eq:gend27}
  f(x,y,z) = (xyz)^{(N-2)/2} \exp[-N(x^2 + y^2 + z^2)/4] 
  \;,\quad
\end{eqnarray}
which for $N\gg1$ is peaked at $(x,y,z)=(1,1,1)$.

\begin{figure}[htb]
  \centering
  \psfig{figure=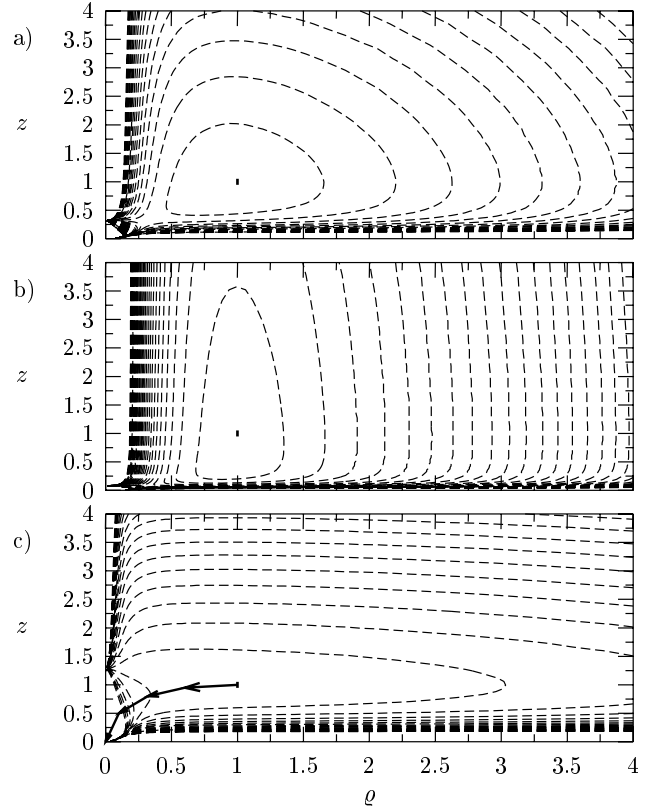,%
    bbllx=4.4cm,bblly=8.5cm,bburx=14.3cm,bbury=20.6cm,angle=0,width=9cm}
  \caption[Contour plots of effective potential for deformed boson system]
  {Contour plots of $u(x,y,z)$, \smarteq{eq:gend17}, with
    $x=y=\varrho$ as a function of $(\varrho,z)$ for
    $s=-0.4\beta^{1/6}$ corresponding to $Na_s/b_\perp=-0.4$ for three
    deformations. The values for the contours change by 2, 2, and 5,
    respectively for a) $\beta=1$ (spherical), b) $\beta=1/16$
    (cigar-shaped or prolate), and c) $\beta=16$ (pancake-shape or
    oblate).  In parts a) and b) the minimum at (1,1) is
      indicated.  In part c) the descending path towards (0,0) is
      indicated.}
  \label{olel3fig1}
\end{figure}

The wave function in \smarteq{eq:gend16} is determined by the
properties of the effective potential $u$.  For axial symmetry around
the $z$ axis the $x$ and $y$ directions cannot be distinguished, that
is when $\gamma=0$ and $\beta=b_\perp^2 / b_z^2$ with $b_\perp^2\equiv
b_xb_y$.  This symmetry amounts to replacing $\rho_x^2$ and $\rho_y^2
$ by $\rho_{\perp}^2/2$ in the equations.  A convenient definition for
this case is $2\varrho^2 \equiv x^2 + y^2$.  Equipotential contours of
$u$ in the $(\varrho,z)$ plane for $\varrho=x=y$ are shown in
\smartfig{olel3fig1} for attractive interactions.  For $a_s<0$ ($s<0$)
there is always a divergence towards $ - \infty$ when $(\varrho,z)
\rightarrow (0,0)$, see \smarteq{eq:gend17}.  However, a stationary
minimum is seen in both \smartfigs{olel3fig1}a (spherical symmetry)
and \ref{olel3fig1}b (prolate) close to $(\varrho,z)=(1,1)$ (indicated
by dots in the figure), whereas this minimum has disappeared for the
oblate system in \smartfig{olel3fig1}c, and no barrier would prevent
contraction (indicated by arrows in the figure).  For weak attraction
a stationary minimum is present for all deformations.

\begin{figure}[htb]
  \centering
  \input{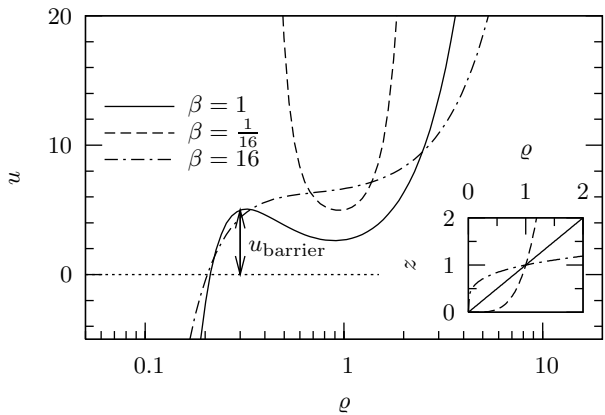}
  \caption[Effective potential along cuts for different deformations]
  {The potential $u(x,y,z)$ for $s=-0.4\beta^{1/6}$ as a function of
    $\varrho=x=y$ along cuts of the $(\varrho,z)$ plane where
    $z=\varrho^{1/\sqrt{\beta}}$.  The height $u\sub{barrier}$ of the
    local maximum at top of the barrier is indicated for the spherical
    case $\beta=1$.  The inset shows corresponding trajectories in the
    $(\varrho,z)$ plane, compare with \smartfig{olel3fig1}, for the
    three deformations.}
  \label{olel3fig2}
\end{figure}

\Smartfig{olel3fig2} shows cuts of the potential $u(\varrho,z)$ along
paths close to the bottom of the valleys (see inset).  The spherical
minimum (full line) is shielded by a relatively small barrier from the
divergence for $\varrho\to0$.  The minimum for the prolate deformation
(dashed curve) is extremely stable although the divergence for
$\varrho\to0$ still exists.  For the oblate deformation (dot-dashed
line) the local minimum has vanished for this attraction strength.

\section{Stability criterion}

\label{sec:stabilitydeformed}

The barrier height depends on the deformation of the external field,
see \smartfigs{olel3fig1} and \ref{olel3fig2}.  Extrema
$(x_0,y_0,z_0)$ of $u$ in \smarteq{eq:gend17} obey the three equations
obtained from
\begin{eqnarray} 
  \frac{b\sub t^2}{b_x^2}(x_0^4-1)
  =
  \sqrt{\frac{2}{\pi}}\frac{sx_0}{y_0z_0}
  \label{eq:extrema}
\end{eqnarray}
and cyclic permutations of $x$, $y$, and $z$.  This can be used to
determine the critical strength $s$ when the local minimum disappears.
The three \smarteqs{eq:extrema} can be reduced to find the parameter
$t\equiv 1/z_0^2-z_0^2$ as a function of $s$ from the equation
\begin{eqnarray}
  &&
  s^2
  =
  \frac\pi2
  \bigg(\frac{b\sub t}{b_z}\bigg)^4
  t^2
  \Bigg[-\frac12q_xt+\sqrt{\frac14q_x^2t^2+1}\Bigg]
  \times
  \label{eq:s_analytical}
  \\
  &&
  \quad
  \Bigg[-\frac12q_yt+\sqrt{\frac14q_y^2t^2+1}\Bigg]
  \Bigg[-\frac12t+\sqrt{\frac14t^2+1}\Bigg]
  \nonumber
  \;,
  \\
  &&
  q_x\equiv\beta+\gamma
  \;,\qquad
  q_y\equiv\beta-\gamma
  \;.
\end{eqnarray}
The maximum value of the right hand side of \smarteq{eq:s_analytical}
is reached when $t$ is the solution to the equation
\begin{eqnarray}
  2
  =
  \frac{q_xt/2}{\sqrt{1+q_x^2t^2/4}}
  +\frac{q_yt/2}{\sqrt{1+q_y^2t^2/4}}
  +\frac{t/2}{\sqrt{1+t^2/4}}
  \label{eq:t-equation}
  \;.
\end{eqnarray}
This solution $t=t\sub{max}$ now from \smarteq{eq:s_analytical} gives
the maximum possible value of $s^2$ when a local minimum of the
effective potential $u$ still is present.  Thus we have obtained the
largest value of $s$ where stable solutions exist.  

\Smarteqs{eq:extrema}-(\ref{eq:t-equation}) are also obtained for the
length parameters $(b_x,b_y,b_z)$ of a deformed harmonic oscillator
ground state wave function by minimizing the expectation value of the
GPE mean-field hamiltonian as formulated by Baym and Pethick
\cite{bay96}.  This relates the present lowest-order hyperspherical
non-correlated results to non-selfconsistent mean-field GPE energies.
However, the present results can be improved by including hyperangles
in the trial wave function.  This enlarges the variational space in
ref.~\cite{bay96} and the solutions are improved.  The comparison to
the self-consistent GPE mean-field results is less direct as the
present results are obtained in a different space and especially the
hyperangular dependence provides a very different structure for the
trial wave function.

\begin{figure}[htb]
  \centering
  \input{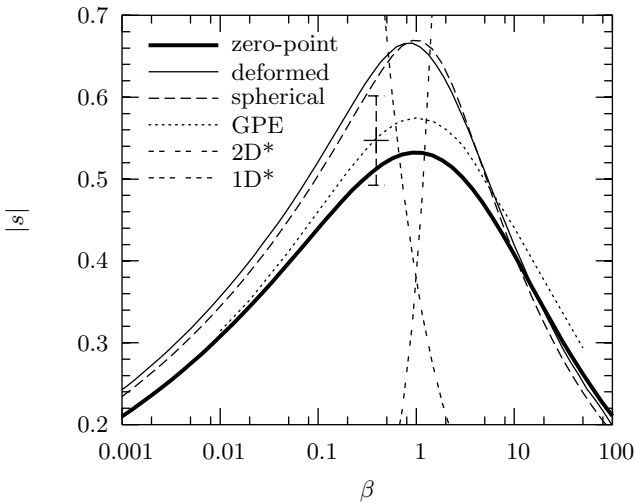}
  \caption[Critical strength as a function of deformation] {The
    critical strength $|s|=N|a_s|/b\sub t$ as a function of the
    deformation $\beta = b_\perp^2/b_z^2$ from the potential in
    \smarteq{eq:gend17} (thin solid line), from \smarteq{eq:gend18}
    (dashed line), and from a mean-field Gross-Pitaevskii computation
    by \etalcite{Gammal}{gam01} (dotted line).  The thick solid line
    is \smarteq{eq:crit_zeropoint} obtained by considering the
    zero-point energy.  The plus is the experimentally measured value
    by \etalcite{Claussen}{cla03} with the 10\% error bars.  Regions
    below curves are considered stable in the separate treatments.
    The double- and triple-dashed lines indicate the effective
    cross-overs to two (2D*) and one (1D*) dimensions from
    \etalcite{G{\"o}rlitz}{gor01b}.}
  \label{olel3fig3}
\end{figure}

The results for axial symmetry ($\gamma=0$) are shown as the thin
solid line in \smartfig{olel3fig3}.  In these units the critical
strength $s$ is largest for a geometry very close to spherical.  Since
$s=Na_s/b\sub t$ and $b\sub t^3=b_xb_yb_z$, this means that at fixed
$b\sub t^3$, or fixed volume, the scattering length can assume the
largest negative value for the spherical trap.
\etalcite{Gammal}{gam01} performed a time-dependent study with the
Gross-Pitaevskii equation (GPE) which resulted in the critical
strengths here shown as the dotted line.  This is in large regions
lower than the present result, which might be due to our neglect of
the quantum effects of the zero-point energy, which is included later
in this paper.  The recent value for the experimental stability region
\cite{cla03} is shown as the plus and agrees with the mean-field
model.  However, the experimental error bars are 10\% and almost
includes the present result.

These results can be compared to an analytical ``spherical''
approximation where the radial motion is described by only $\rho$
while the deformed external field remains the same.  The effective
radial potential $U$ is then obtained by adding centrifugal barrier
and the contributions from zero-range interaction and the external
field.  The angular average replaces each of the three components
$\rho_q^2$ and $R_q^2$ by $\rho^2/3$ and $R^2/3$, where $\boldsymbol
R$ is the center-of-mass coordinates, i.e.
\begin{eqnarray}
  &&
  \sum_{i=1}^N \big\langle V\sub{trap}(\boldsymbol r_i) \big\rangle_{\Omega}
  =
  \frac12m \frac{\omega_x^2+\omega_y^2+\omega_z^2}{3} 
  \big(\rho^2 + N R^2\big)
  \;,\qquad
  \\
  &&
  \frac{2m \hat V}{\hbar^2}
  =
  8\pi a_s
  \sum_{i<j}^N 
  \big\langle \delta(\boldsymbol r_{ij}) \big\rangle_\Omega 
  = 
  \frac{3}{2}
  \sqrt{\frac{3}{\pi}}
  N^{7/2}
  \frac{a_s}{\rho^3}
  \label{eq:vdelta_copy1}
  \;,
  \\
  &&
  \frac{2mU(\rho)}{\hbar^2}  
  =
  \frac{3}{2}
  \sqrt{\frac{3}{\pi}}
  N^{7/2}
  \frac{a_s}{\rho^3} + 
  \frac{9N^2}{4\rho^2}+
  \frac{\rho^2}{l_2^4} 
  \;,
  \label{eq:ueff_deformed}
\end{eqnarray}
where $3l_2^{-4} \equiv b_x^{-4}+b_y^{-4}+b_z^{-4}$. The stability
condition becomes
\begin{eqnarray}
  &&
  \frac{N|a_s|}{b\sub t} <  k(\beta,\gamma)
  \;,\qquad
  k(1,0)=\frac{2\sqrt{2\pi}}{5^{5/4}} \simeq 0.67   
  \label{eq:gend18}
  \;,\;
  \\
  &&
  k(\beta,\gamma)
  =
  k(1,0)
  \sqrt[4]{\frac{3(\beta^2-\gamma^2)^{4/3}}
    {2\beta^2+2\gamma^2+(\beta^2-\gamma^2)^2}}
  \label{eq:gend29}
  \;.
\end{eqnarray}

The spherical limit corresponds to $\gamma=0$ and $\beta=1$ where the
barrier is present when $|s|= N |a_s| / b\sub t < 0.67$.  The result
for a cylinder (only $\gamma=0$) is shown as the dashed line in
\smartfig{olel3fig3} and is noticably different from, but numerically
almost coincides with, the ``deformed'' treatment, thin solid line.
An extreme oblate deformation corresponds to the two-dimensional limit
where $b_z\ll b_\perp$ and $\beta\to\infty$.  Here \smarteq{eq:gend29}
yields the critical strength $k \simeq 0.4 \sqrt[4]{0.6} \sqrt{\pi/2}
\beta^{-1/3}$.  As seen from the contour plot in
\smartfig{olel3fig1}c, the motion is now almost confined at $z=1$.
From $x=y=\varrho$ we see that $u(\varrho,\varrho,1)$ only has a local
minimum when $|s|<\sqrt{\pi/2}\beta^{-1/3}$, which is larger than the
value where the $z$ motion is not fixed.  This is reasonable since
more degrees of freedom in the model lowers the energy.
\Smarteq{eq:s_analytical} and Baym and Pethick \cite{bay96} give the
same value in this limit, the latter obtained with a variational study
of the GPE provided that the variational width in the axial direction
does not change due to the interactions.  This is identical to the
criterion from studying the potential $u(\varrho,\varrho,1)$,
i.e.~consistent with the fixed value $z=1$.

Analogously, in the extreme prolate limit (one-dimensional) where
$\beta\to0$, \smarteq{eq:gend29} yields the critical strength
$k\simeq0.25\sqrt[4]{1.25}\sqrt{\pi}\beta^{1/6}$.  From
\smarteq{eq:s_analytical} we obtain in this limit $k\simeq
3^{-3/4}\sqrt{\pi}\beta^{1/6}$, which has the same deformation
dependence, but is a factor $\sim1.7$ larger than the value from
\smarteq{eq:gend29}.  However, fixing $x=y=1$ in \smarteq{eq:gend17}
yields no critical strength since $u(1,1,z)$ always has a global
minimum.  Therefore, the other degrees of freedom are essential in
this prolate limit.

A better stability criterion can be obtained by considering the
ground-state energy $E_0$ of the boson system, which in the
non-interacting case is $E_0=\hbar(\omega_x+\omega_y+\omega_z)(N-1)/2$
where the center-of-mass energy is subtracted.  The system is unstable
when this energy is larger than the barrier height $U\sub{barrier}$ of
the hyperradial potential $U(\rho)$; see the indication in
\smartfig{olel3fig2} of the corresponding height $u\sub{barrier}$ for
the reduced potential $u$.  With this condition the criterion of
stability is
\begin{eqnarray}
  \frac{N|a_s|}{b\sub t}
  <
  \frac12
  \sqrt{\frac\pi3}\frac{l_1}{b\sub t}
  \sqrt{1+\frac{1}{12}\frac{l_1^4}{l_2^4}}
  \;,
  \label{eq:crit_zeropoint}
\end{eqnarray}
where $3l_1^{-2} \equiv b_x^{-2}+b_y^{-2}+b_z^{-2}$.  This is seen in
\smartfig{olel3fig3} (thick solid line) to be below the GPE
calculations \cite{gam01}.  The improvement is here substantial
compared to when the zero-point energy is neglected.  In particular,
for the spherical case we get $N|a_s|/b\sub t\simeq0.53$ instead of
$N|a_s|/b\sub t\simeq0.67$.  This is within the 10\% error bars of the
experimental value.  The estimate of \smarteq{eq:crit_zeropoint}
describes the stability problem better than
\twoeqs{eq:gend18}{eq:gend29} since it includes the quantum effect due
to the zero-point energy.  An improvement on this would be to solve
the hyperradial problem with interaction effects, which would lower
the zero-point energy.  The critical value would consequently
increase, making the present value a lower bound.

A recent variational Monte Carlo investigation of the stability
criterion in elongated, almost one-dimensional, traps yielded the
stability criterion $n\sub{1D}a\sub{1D} \lesssim 0.35$ \cite{ast03},
where $n\sub{1D}\sim N/b_z$ is the density in one dimension and
$a\sub{1D} = -b_\perp(b_\perp/a_s-1.0326)$.
\Smarteq{eq:crit_zeropoint} can in the one-dimensional limit be
written as $N|a_s|/b_\perp \lesssim 0.66$.  The deviation between the
two results might be due to our use of a three-dimensional zero-range
interaction in this non-correlated model, whereas
\etalcite{Astrakharchik}{ast03} used a one-dimensional model with a
zero-range interaction with coupling strength proportional to
$1/a\sub{1D}$ as well as a full 3D correlated model with hard-sphere
or finite-range potentials.  An effective potential analogous to
$\delta(x)/a\sub{1D}$ in the general case with intermediate
deformations would be a rewarding goal.

According to \etalcite{G\"orlitz}{gor01b} the interaction energy is
smaller than the energy in the tightly-confining dimension when
$|s|\le\sqrt{32/225}\beta^{-5/6}$ for the 1D limit and when $|s|\le
\sqrt{32/225}\beta^{5/3}$ for the 2D limit.  These cross-overs are 
indicated by double-dashed (two-dimensional) and triple-dashed
(one-dimensional) lines in \smartfig{olel3fig3}.  Since the critical
region in each limit is below the relevant cross-over, stable and
strongly deformed systems can be regarded as effectively one- or
two-dimensional in the sense of these energy relations.

\section{Effective dimension}

\label{sec:effdim}

The deformation of the external field effectively changes the
dimension $d$ of the space where the particles move.  The field
changes continuously and $d$ could correspondingly vary from three to
either two or one.  In order to arrive at such a description, we aim
at an effective $d$-dimensional Hamiltonian analogous to
\smarteq{eq:gend1} with only one radial variable $\rho$, a
deformation-dependent dimension $d$, and an effective trap length
$b_d$, i.e.
\begin{eqnarray}
  \frac{2m\hat H_d}{\hbar^2}
  =
  -\frac{1}{\rho^{d(N-1)-1}}
  \frac{\partial}{\partial\rho}
  \rho^{d(N-1)-1}
  \frac{\partial}{\partial\rho}
  +\frac{\rho^2}{b_d^4} + \frac{2m\hat V}{\hbar^2}
  \label{eq:gend6}
  \;,\quad
\end{eqnarray}
where $\hat V$ represents all particle interactions in $d$ dimensions.
The requirement is that the Schr\"odinger equation $\hat H_d G_d = E_d
G_d$ with $d$-dimensional eigenfunction $G_d$ and eigenvalue $E_d$ is
obeyed, at least on average, i.e.
\begin{eqnarray}
  \int d\rho\;\rho^{d(N-1)-1}
  G_d^*(\rho)
  \big(\hat H_d-E_d\big)
  G_d(\rho)
  =
  0
  \;.
  \label{eq:gend7}
\end{eqnarray}

The lowest free solution, that is with $\hat V=0$, is given by
\smarteq{eq:gend27}.  In the cylindrical case we can relate the
$d$-dimensional function $G_d$ to this by performing the average with
respect to the angle $\theta$ in the parametrization $(\rho_\perp,
\rho_z)=\rho( \sin \theta, \cos \theta)$.  With inclusion of the
corresponding volume elements this leads to
\begin{eqnarray}
  &&
  \rho^{d(N-1)-1}|G_d(\rho)|^2 
  =
  \rho^{3(N-1)-1}
  \times
  \nonumber\\
  &&\qquad
  \int_0^\pi d\theta \cos^{N-2}\theta\sin^{2N-3}\theta 
  |F(\rho,\theta)|^2
  \label{eq:gend40}
  \;,
\end{eqnarray}
where $F(\rho,\theta)$ can be obtained by rewriting
\twoeqs{eq:gend27}{eq:fFtrans}.

The characteristic energy and length can be defined by
\begin{eqnarray}
  E_d =
  \frac{d\hbar^2}{2mb_d^2}(N-1)
  \;,\qquad
  db_d^2=2b_\perp^2+b_z^2
  \;,
  \label{eq:gend3}
\end{eqnarray}
which clearly is correct in the three limits, i.e.~spherical: $d=3$
and $b_d = b_z = b_\perp$, two-dimensional: $d=2$ and $b_\perp\gg
b_z$, and one-dimensional: $d=1$ and $b_z\gg b_\perp$.

\begin{figure}[htb]
  \centering
  \input{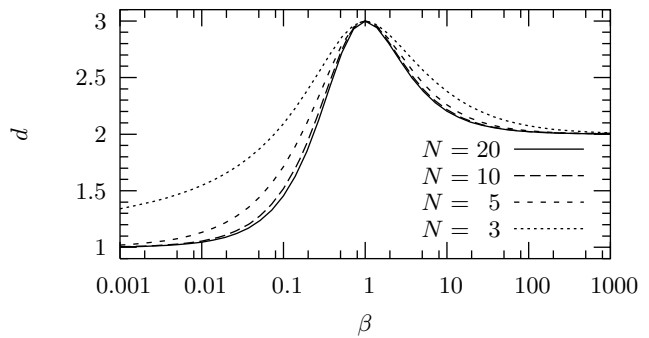}
  \caption[Effective dimension as a function of deformation] 
  {The effective dimension $d$ obtained as a function of the
    deformation parameter $\beta=b_\perp^2/b_z^2$.  Curves for larger
    $N$ are very close to that for $N=20$.}
  \label{olel3fig4}
\end{figure}

In general it is not possible to find one $\rho$-independent set of
constants ($E_d,b_d,d$) such that $\hat H_d G_d = E_d G_d$.  Instead
we insist on the average condition in \smarteq{eq:gend7} with $G_d$
and $E_d$ from \twoeqs{eq:gend40}{eq:gend3}.  The result for axial
geometry is a second-degree equation in $d$ with one physically
meaningful root.  The results for various $N$ values are shown in
\smartfig{olel3fig4}.
The effective dimension depends on $N$ for relatively small particle
numbers.  When $N>20$, the curve is essentially fixed.  Furthermore,
the asymptotic values of both $d=1$ (small $\beta$) and $d=2$ (large
$\beta$) are reached faster for larger $N$ since many particles feel
the geometric confinement stronger than few particles.  Since these
effective dimensions are obtained as average values over $\rho$, the
system might look spherical at large distances and strongly deformed
at small distances, on average resulting in the curves in
\smartfig{olel3fig4}.

\section{Deformation-dependent interactions}

\label{sec:effint}

The effective dimension for the non-interacting system possibly
changes when interactions are included.  The steps of the previous
section should in principle be repeated with the interactions.
However, this would be complicated and miss the goal which is a simple
effective Hamiltonian with a renormalized interaction in lower
dimension, see analogies in the references \cite{ols98,pet00,lee02}.

We therefore start out with a two-body contact interaction with a
coupling strength which is modified due to the deformation.  This is
in line with usual renormalizations due to density-dependent effects
\cite{lee57b,bra02b}; see also a proposed modification due to the
inclusion of two-body correlations in the reference \cite{sor04}.  So
we write a $d$-dimensional zero-range interaction with a
dimension-dependent coupling strength $g(d)$ as
\begin{eqnarray}
  V_d(r_{ij})
  =
  g(d) \delta^{(d)}(r_{ij})
  \;,\qquad
  g(3)=\frac{\hbar^2a_s}{m}
  \label{eq:gend36}
  \;,
\end{eqnarray}
where this ``$d$-dimensional $\delta$ function'' is defined by
$\delta^{(d)}(r)=0$ for $r\ne0$ and $\int_0^\infty
dr\;r^{d-1}\delta^{(d)}(r) = 1$.  The distance between two particles,
e.g.~particle 1 and 2, is in hyperspherical coordinates defined by
$r_{12}=\sqrt{2} \rho \sin \alpha$, where the angle $\alpha$ enters
the angular volume element as
\begin{eqnarray}
  d\Omega_\alpha
  = d\alpha\sin^{d-1}\alpha \cos^{d(N-2)-1}\alpha
  \;.
\end{eqnarray}
This is valid at least for $d=1,2,3$.  The effective interaction $\hat V$
in \smarteq{eq:gend6} is for $N\gg1$ then given by the average
over all coordinates except $\rho$:
\begin{eqnarray}
  \hat V
  &=&
  \frac{N^2}{2}
  \frac{\int_0^{\pi/2}d\Omega_\alpha \;
    V_d(\sqrt2\rho\sin\alpha)}
  {\int_0^{\pi/2}d\Omega_\alpha}
  \nonumber\\
  &=&
  \frac{\hbar^2}{2m}
  \frac{2N^2(Nd/4)^{d/2}}{\Gamma(d/2)}
  \frac{a_s}{\rho^d}
  \frac{g(d)}{g(3)}
  \label{eq:hatvd}
  \;.
\end{eqnarray}
However, this does not yield instability for $d<2$ since the power $d$
in $\rho^{-d}$ is smaller than two.

\begin{figure}[htb]
  \centering
  \input{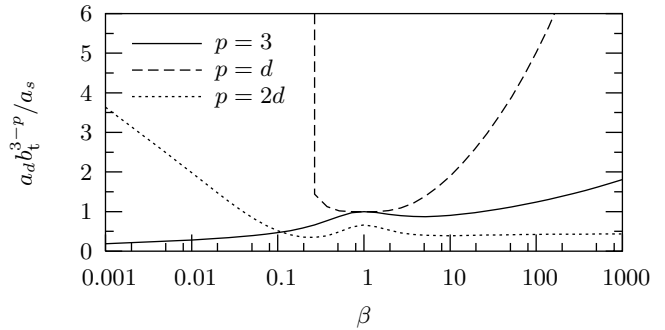}
  \caption
  [Effective interaction strength as a function of deformation] {The
    effective interaction strength $a_d$ from \smarteq{eq:def_intstr}
    obtained as a function of the deformation parameter
    $\beta=b_\perp^2/b_z^2$ in the large-$N$ limit, i.e.~the
    connection between deformation and effective dimension obtained
    from the calculation for $N=20$ is used for this illustration.
    The vertical divergence of the dashed line indicates the
    inadequacy of the corresponding method when $d<2$.}
  \label{fig:def_intstr}
\end{figure}

We therefore pursue another approach.  Inspired by the forms of
\twoeqs{eq:vdelta_copy1}{eq:hatvd}, we write $\hat V$ as
\begin{eqnarray}
  \hat V
  =
  \frac{\hbar^2}{2m}
  \frac{2N^2(Nd/4)^{p/2}}{\Gamma(d/2)}
  \frac{a_d}{\rho^p}
  \;,\qquad
  a_3=a_s
  \;,
\end{eqnarray}
which with $a_3=a_s$ coincides with the result for $d=3$ if we choose
$p=3$.  The effective potential $U_{d}$ in the $d$-dimensional
Schr\"odinger equation corresponding to \smarteq{eq:gend6} is
then
\begin{eqnarray}
  \frac{2mU_{d}(\rho)}{\hbar^2}  =
  \frac{2N^2(Nd/4)^{p/2}}{\Gamma(d/2)}
  \frac{a_d}{\rho^p}
  + 
  \frac{d^2N^2}{4\rho^2}+
  \frac{\rho^2}{b_d^4} 
  \;.
\end{eqnarray}
For $p<2$ this potential always has a global minimum and thus no
collapse is present.  For $p>2$ there is always divergence to
$-\infty$ when $\rho\to0$.  For weak attraction, i.e.~small $|a_d|$,
there is a local minimum.  This disappears at larger $|a_d|$ when
\begin{eqnarray}
  \frac{N|a_d|}{b\sub t}
  >
  \frac{b_d^{p-2}}{b\sub t}
  \frac{2^{1+p/2}d(p-2)^{(p-2)/4}\Gamma(d/2)}{p(p+2)^{(p+2)/4}}
  \;.
  \label{eq:stability_effective_interaction}
\end{eqnarray}
The criterion in \smarteq{eq:gend18} was also obtained by estimating
when the critical point vanished.  \Smarteq{eq:gend18} is valid for
all deformations, i.e.~any $d$.  In order to be able to compare
\twoeqs{eq:gend18}{eq:stability_effective_interaction}, we therefore
choose $p>2$ such that \smarteq{eq:stability_effective_interaction}
always is applicable.  When
\twoeqs{eq:stability_effective_interaction}{eq:gend18} agree, the
effective interaction strength $a_d$ is given by
\begin{eqnarray}
  \frac{a_d}{a_s}
  =
  \frac{b_d^{p-2}}{b\sub t}
  \frac{2^{(p-1)/2}\Gamma(\frac{d}{2})5^{5/4}(p-2)^{(p-2)/4}}
  {\sqrt{\pi}(p+2)^{(p+2)/4}\beta^{1/6}p/d}
  \sqrt[4]{\frac{2+\beta^2}{3}}
  \label{eq:def_intstr}
  \;.\quad
\end{eqnarray}
This effective interaction strength is in \smartfig{fig:def_intstr}
shown as a function of the deformation for various choices of the
power $p$.  The solid line shows the result for $p=3$, which is known
to be correct for $\beta=1$ ($d=3$).  Similarly the dashed line shows
the result with $p=d$, which does not work for $d<2$
($\beta\lesssim0.2$).
Since the effective coupling strength depends strongly on the power
$p$, we need further information about how the interactions enter the
effective potential.

An extreme deformation might lead to effectively one-dimensional or
two-dimensional properties.  Pitaevskii and Stringari \cite{pit03}
collected results for the effective coupling strength in two
dimensions that yields $g(2)=\sqrt{8\pi}\hbar^2a_s/(mb_z)$, whereas
the result from \smarteq{eq:def_intstr} in that limit is larger by the
factor $5^{5/4}/4\simeq 1.9$.  Even though the results differ by a
factor close to two, the right combination of lengths shows that we
have incorporated the degrees of freedom in the correct manner.  This
was also the case in the previous comparison of the stability
criterion with the one obtained by Baym and Pethick \cite{bay96}.
However, as was also mentioned by Pitaevskii and Stringari
\cite{pit03}, in the low-density limit in two dimensions the coupling
constant becomes density-dependent, which is beyond the present model
where correlations are neglected.

Since $p=d$ for $d=1$ does not agree with a meaningful interpretation
of the stability criterion, a one-dimensional system needs a different
treatment.

\section{Discussion}

In conclusion, the hyperspherical method with a non-correlated
approach yielded stability criteria as a function of the deformation
of the external field.  For constant volume the highest stability was
found for spherical traps.  Effective dimensions $d$ continuously
varying between 1 and 3 were calculated as a function of the
deformation.  The system can be described by a $d$-dimensional
effective radial potential with a $d$-dimensional effective
interaction.  However, this does not have an unambigious form.
Applications to restricted geometries become simpler, where the
obtained two-dimensional coupling strength compares reasonably with a
coupling strength obtained by an axial average of a three-dimensional
contact interaction.  For the one-dimensional case an effective
coupling strength was not obtained.

A previous approach to a $d=1$ treatment by \etalcite{Gammal}{gam00}
shows that a three-body contact interaction is necessary for the GPE
to produce collapse in one spatial dimension.  In the present
framework a three-body contact interaction for a constant angular wave
function produces a hyperradial potential proportional to $\rho^{-2d}$
which for any $d\ge1$ leads to instability if the three-body coupling
strength is sufficiently negative.  The dotted line in
\smartfig{fig:def_intstr} shows the effective coupling strength for
$p=2d$, corresponding to this three-body zero-range interaction.

According to \etalcite{Astrakharchik}{ast03,ast03b} a Jastrow ansatz
for a correlated wave function and inclusion of two-body interactions
lead to collapse in one spatial dimension.  According to Faddeev
calculations with the two-body correlated model presented in reference
\cite{sor02b}, a two-body interaction and inclusion of only two-body
correlations in one spatial dimension do not lead to collapse.  It
seems that at least three-body correlations or three-body interactions
are necessary in order to achieve a realistic description of collapse
in one dimension.

\bibliographystyle{prsty}
\bibliography{../bibliografi.bib}

\end{document}